\documentclass[twocolumn]{jpsj3}
\usepackage{txfonts}
\usepackage{bm}

\title{Complex-Stripe Phases Induced by Staggered Rashba Spin-Orbit Coupling}

\author{Tomohiro Yoshida$^1$, Manfred Sigrist$^2$, and Youichi Yanase$^1$}
\inst{$^1$Department of Physics, Niigata University, Niigata 950-2181, Japan \\
$^2$Theoretische Physik, ETH-Zurich, 8093 Zurich, Switzerland}

\abst{
We study superconducting phases in a quasi-two-dimensional multilayer system without local inversion symmetry.
Broken local inversion symmetry induces layer-dependent Rashba-type spin-orbit couplings.
We find that a complex-stripe phase, which is the intermediate phase between the Fulde-Ferrell (FF) phase and Larkin-Ovchinnikov (LO) phase, is realized in the magnetic field applied parallel to the layers. 
A crossover from the FF phase to the LO phase appears by tuning temperature and magnetic field. 
We show the local density of states that characterizes the complex-stripe phase. 
As a possible realization of the complex-stripe phase, we discuss the artificial superlattices of CeCoIn$_5$.
}

\kword{locally non-centrosymmetric superconductors, staggered Rashba spin-orbit coupling, complex-stripe phase, phase diagram}

\begin{document}

\maketitle

\section{Introduction}\label{sec:sec1}

Since the discovery of the non-centrosymmetric superconductor CePt$_3$Si,~\cite{PhysRevLett.92.027003} numerous studies of non-centrosymmetric superconductivity have been conducted, both theoretically and experimentally, elucidating unusual properties.~\cite{NCSC}
Examples are the mixing of spin-singlet and triplet Cooper pairings,~\cite{SovPhysJETP.68.1244,PhysRevLett.87.037004} magnetoelectric effect,~\cite{PhysRevB.72.172501,PhysRevB.65.144508,PhysRevB.72.024515,JPSJ.76.034712} helical superconducting phase,~\cite{SovPhysJETP.68.1244,JETPLett.78.637,PhysRevB.70.104521,PhysRevLett.94.137002,PhysRevB.75.064511,PhysRevB.76.014522,JPSJ.76.124709,PhysRevB.78.220508} 
and topological superconductivity.~\cite{PhysRevB.79.094504,PhysRevLett.103.020401}
These unconventional features result from an antisymmetric spin-orbit coupling, which can be classified in various categories such as Rashba-type, Dresselhaus-type, and cubic-type. Also, in systems designed for ultracold atomic gases antisymmetric spin-orbit coupling has become a technically feasible ingredient recently.~\cite{Nature.471.83}

In a number of theoretical studies, unusual features are predicted for systems that conserve inversion symmetry globally but not locally.~\cite{JPSJ.79.084701,PhysRevB.84.184533,JPSJ.81.034702,PhysRevB.85.220505,PhysRevB.86.100507,PhysRevB.86.134514} 
For instance, a pair-density wave (PDW) state could arise from the 
staggered antisymmetric spin-orbit coupling inherent to ``local non-centrosymmetricity'' in materials.~\cite{PhysRevB.86.134514} 
These works discuss multilayer superconductors, such as artificial superlattices of CeCoIn$_5$,~\cite{NatPhys.7.849} multilayer high-$T_{\rm c}$ cuprates,~\cite{PhysRevLett.96.087001,JPSJ.80.043706} and SrPtAs~\cite{JPSJ.80.055002} as well as disordered phases of CePt$_3$Si~\cite{JPSJ.78.014705} and Sr$_2$RuO$_4$.~\cite{PhysRevB.77.214511,1742-6596-150-5-052113,0295-5075-83-2-27007}

A spatially modulated superconducting state [e.g., Fulde-Ferrell-Larkin-Ovchinnikov (FFLO) state] can be induced to circumvent paramagnetic limiting 
in a magnetic field.~\cite{PhysRev.135.A550,ZhEkspTeorFiz.47.1136}
While a Larkin-Ovchinnikov (LO) state, characterized by the magnitude modulation of the superconducting order parameter, 
is expected in centrosymmetric superconductors,~\cite{JPSJ.76.051005} a Fulde-Ferrell (FF) state, a ``helical phase'' with an essentially constant order parameter magnitude,~\cite{SovPhysJETP.68.1244,JETPLett.78.637,PhysRevB.70.104521,PhysRevLett.94.137002,PhysRevB.75.064511,PhysRevB.76.014522,JPSJ.76.124709,PhysRevB.78.220508} could appear in non-centrosymmetric superconductors exposed to a magnetic field. 
Although experimental indications of the realization of an LO phase have been found for the heavy-fermion superconductor CeCoIn$_5$,~\cite{Nature.425.51,PhysRevLett.91.187004} organic superconductors,~\cite{PhysRevLett.97.157001,PhysRevLett.99.187002,PhysRevB.83.064506} ultracold fermion gases with a population imbalance,~\cite{Nature.467.567} and nuclear matter,~\cite{RevModPhys.76.263} there has been no experimental evidence of the helical phase in non-centrosymmetric superconductors so far. 
In this paper, we show that a complex-stripe phase, which can be viewed as intermediate between FF and LO states, may exist in locally non-centrosymmetric multilayers. As shown below, such a phase may cover a large region in the $T$-$H$ phase diagram of locally non-centrosymmetric superconductors, in contrast to a rather tiny region for the standard LO phase.~\cite{JPSJ.80.123706} 
In contrast to the helical phase in non-centrosymmetric superconductors, a spatially inhomogeneous complex-stripe phase could be observed in the experiments with high spatial resolution, such as scanning tunneling microscopy (STM) and NMR.

We start this paper with the basic concept of the complex-stripe phase in Sect.~\ref{sec:sec2}, followed by
the introduction of a minimal Hamiltonian for a locally non-centrosymmetric multilayer superconductors in Sect.~\ref{sec:sec3}. 
In Sect.~\ref{ssec:ssec4-1}, we investigate the $T$-$H$ phase diagram for the bilayer system and show that the complex-stripe phase is stabilized 
by staggered Rashba spin-orbit coupling. For comparison with possible experiments we explore the local density of states in the complex-stripe phase in Sect.~\ref{ssec:ssec4-2}. We also study the complex-stripe phase in a trilayer system in Sect.~\ref{sec:sec5}. Finally, we discuss the possibility of the formation of the complex-stripe phase in artificial superlattices of CeCoIn$_5$ in Sect.~\ref{sec:sec6}.

\section{Complex-Stripe Phase}\label{sec:sec2}

We start our discussion by introducing the concept of the complex-stripe phase.
For this purpose, we restrict ourselves to the simplest nontrivial case, the bilayer system. 
\begin{figure}[htbp]
 \includegraphics[width=85mm]{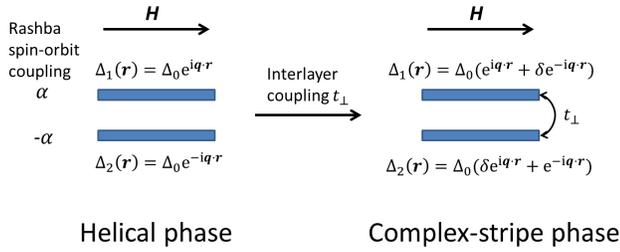}
 \caption{(Color online) Schematic figures of the complex-stripe phase. The thick bars describe the two-dimensional conducting planes.
   The structures of the staggered Rashba spin-orbit coupling are shown on the left-hand side of the figure.
   The order parameters of the upper and lower layers $\Delta_{1,2}({\bm r})$ are shown on the top or at the bottom of the bars.
   For the left (right)-hand figure, a bilayer system without (with) interlayer coupling $t_\perp$ is assumed.
   See text for details.}
 \label{fig:fig1}
\end{figure} 
As shown in Fig.~\ref{fig:fig1}, the local inversion symmetry is broken for each layer yielding an in-plane Rashba spin-orbit coupling. 
Global inversion symmetry requires that the coupling constants of the Rashba spin-orbit coupling $\alpha_m$ are layer-dependent and antisymmetric with respect to reflection at the center of the multilayer structure and $(\alpha_1,\alpha_2)=(\alpha,-\alpha)$ for bilayers.
If we neglect the interlayer coupling and take the layers as completely decoupled, then a magnetic field would induce a helical phase 
as in non-centrosymmetric superconductors. Because of the opposite sign of 
Rashba spin-orbit coupling, the wave vector ${\bm q}$ of the order parameter would have opposite signs in the two layers. 
Thus, the order parameter can be represented as $\Delta_1({\bm r})=\Delta_0 {\rm e}^{{\rm i}{\bm q}\cdot{\bm r}} $ on the upper layer and as
$\Delta_2({\bm r})=\Delta_0 {\rm e}^{-{\rm i}{\bm q}\cdot{\bm r}}$ on the lower layer 
(see the left-hand side of Fig.~\ref{fig:fig1}). 
Now, we turn the interlayer coupling $t_\perp$ on and find that the complex order parameters of two layers mix with each other leading to $\Delta_1({\bm r})=\Delta_0({\rm e}^{{\rm i}{\bm q}\cdot{\bm r}}+\delta {\rm e}^{-{\rm i}{\bm q}\cdot{\bm r}})$ and $\Delta_2({\bm r})=\Delta_0(\delta {\rm e}^{{\rm i}{\bm q}\cdot{\bm r}} +  {\rm e}^{-{\rm i}{\bm q}\cdot{\bm r}})$ (see the right-hand side of Fig.~\ref{fig:fig1}). This new order parameter structure corresponds to a complex phase with a stripe structure, i.e., the complex-stripe phase. From its structure it can be regarded as intermediate between the FF state, which has the order parameter $\Delta({\bm r})=\Delta_0{\rm e}^{{\rm i}{\bm q}\cdot{\bm r}}$, and the LO state with $\Delta({\bm r})=\Delta_0\cos({\bm q}\cdot{\bm r})$.
Obviously, we have a modulation of the order parameter magnitude as well as a helical modulation, e.g., $ \Delta_1({\bm r}) = \Delta_0 {\rm e}^{{\rm i}{\bm q}\cdot{\bm r}} ( 1 + \delta {\rm e}^{-2{\rm i}{\bm q}\cdot{\bm r}} ) \approx \Delta_0 {\rm e}^{{\rm i}{\bm q}\cdot{\bm r}} [ 1 + \delta^2 +  2\delta \cos(2{\bm q}\cdot{\bm r}) ]^{1/2} $. 

It is important to note here that the helical phase with a wave vector ${\bm q} \sim2\mu_{\rm B}  (\hat{z} \times {\bm H}) \alpha/v_{\rm F}E_{\rm F}$ in a homogeneous non-centrosymmetric superconductor cannot be observed directly owing to gauge invariance.~\cite{PhysRevB.78.220508} This is different for the complex-stripe phase where different wave vectors on the two layers cannot be removed by gauge transformation and would yield observable results similar to the situation discussed recently in Ref.~\citen{PhysRevLett.109.237007}.

\section{Model and BdG Equations}\label{sec:sec3}

We now turn to the two-dimensional multilayer model with staggered Rashba spin-orbit coupling, which we formulate as follows:
\begin{eqnarray}
\mathcal{H}&=&-t\sum_{\langle{\bm i},{\bm j}\rangle,s,m}c_{{\bm i}s m}^\dagger c_{{\bm j}s m}-\mu\sum_{{\bm i},s,m}c_{{\bm i}s m}^\dagger c_{{\bm i}s m} \nonumber \\
&&-\mu_{{\rm B}}\sum_{{\bm i},s,s',m}{\bm H}\cdot{\bm \sigma}_{ss'}c_{{\bm i}s m}^\dagger c_{{\bm i}s'm} \nonumber \\
&&+t_{\perp}\sum_{{\bm i},s,\langle m,m'\rangle}c_{{\bm i}sm}^\dagger c_{{\bm i}s m'} \nonumber \\
&&-{\rm i}\frac{1}{2}\sum_{\langle{\bm i},{\bm j}\rangle,s,s',m}\alpha_m({\bm \sigma}_{ss'}\times \hat{{\bm r}}_{ij})_z c_{{\bm i}sm}^\dagger c_{{\bm j}s'm} \nonumber \\
&&-V\sum_{{\bm i},m}c_{{\bm i}\uparrow m}^\dagger c_{{\bm i}\downarrow m}^\dagger c_{{\bm i}\downarrow m}c_{{\bm i}\uparrow m},
\label{eq:eq1}
\end{eqnarray}
where $c_{{\bm i}sm}^\dagger$ ($c_{{\bm i}sm}$) is the creation (annihilation) operator for an electron with a spin $s$ on a site ${\bm i}=(i_x,i_y)$ and a layer $m$. 
$\hat{{\bm r}}_{ij}$ is a unit vector connecting  ${\bm i}$ with the nearest-neighbor site ${\bm j}$.
The symbol $\langle {\bm i},{\bm j}\rangle$ ($\langle m,m'\rangle$) denotes the summation over nearest-neighbor sites (layers).

For our model calculation, we choose the intralayer hopping $t$ as the energy unit and assume a small interlayer coupling $t_\perp/t=0.1$, and chemical potential $\mu/t=2$. 
The magnetic field ${\bm H}$ is applied parallel to the conducting plane ${\bm H}=(H,0,0)$ parallel to the $x$-axis. We neglect here 
the orbital coupling of the magnetic field, assuming the effect of orbital pair-breaking as small, since we focus on quasi-two-dimensional electron systems such as multilayer high-$T_{\rm c}$ cuprates and artificial superlattices of CeCoIn$_5$. 
As mentioned earlier, we choose the coupling constants of the spin-orbit coupling to be equal in magnitude but opposite in sign on the two layers, namely, $(\alpha_1,\alpha_2)=(\alpha,-\alpha)$ for the bilayer system. For the trilayer system, however, Rashba spin-orbit coupling vanishes on the 
center layer by symmetry such that $(\alpha_1,\alpha_2,\alpha_3)=(\alpha,0,-\alpha)$. 
The last term in Eq.~(\ref{eq:eq1}) denotes the intralayer pairing interaction for spin-singlet $s$-wave superconductivity. 
The choice of pairing symmetry, in the spin-singlet channel, is irrelevant for the qualitative discussion of the behavior in an in-plane magnetic field.
Thus, we prefer the numerically simpler case of $s$-wave pairing over $d$-wave pairing, although the latter would be relevant for
CeCoIn$_5$.

We analyze this model by solving the Bogoliubov$-$de Gennes (BdG) equations.
First, we perform the mean-field approximation of the pairing interaction, which leads to the mean-field Hamiltonian $\mathcal{H}_{\rm MF}$. 
For the magnetic field along the $x$-axis, the spatial modulation of the order parameter appears along the $y$-axis.~\cite{PhysRevB.75.064511,PhysRevB.76.014522} 
Because the order parameter is uniform along the $x$-axis, we describe the mean-field Hamiltonian $\mathcal{H}_{\rm MF}$ using the momentum representation along the $x$-direction. 
Then, the mean field Hamiltonian $\mathcal{H}_{\rm MF}$ is obtained as
\begin{eqnarray}
\mathcal{H}_{\rm MF}&=&\sum_{k_x,i_y,j_y,s,m}
[-t(\delta_{i_y+1,j_y}+\delta_{i_y-1,j_y}) \nonumber \\
&&+(-2t\cos k_x-\mu)\delta_{i_y,j_y}]c^\dagger_{i_y sm}(k_x) c_{j_y sm}(k_x) \nonumber \\
&&-\mu_{\rm B}H\sum_{k_x,i_y,s,m}c^\dagger_{i_y s m}(k_x)c_{i_y\bar{s} m}(k_x) \nonumber \\
&&+t_\perp\sum_{k_x,i_y,s,\langle m,m'\rangle}c^\dagger_{i_y s m}(k_x) c_{i_y s m'}(k_x) \nonumber \\
&&+\sum_{k_x,i_y,j_y,m}\biggl[-{\rm i}\frac{\alpha_m}{2}(\delta_{i_y+1,j_y}-\delta_{i_y-1,j_y} \nonumber \\
&&+2\sin k_x \delta_{i_y,j_y})c^\dagger_{i_y \uparrow m}(k_x)c_{j_y \downarrow m}(k_x)+{\rm H.c.}\biggr] \nonumber \\
&&+\sum_{k_x,i_y,m}[\Delta_{i_y,m}c^\dagger_{i_y \uparrow m}(k_x)c^\dagger_{i_y \downarrow m}(-k_x)+{\rm H.c.}] \nonumber \\
&&+\frac{N_x}{V}\sum_{i_y,m}|\Delta_{i_y,m}|^2,
\label{eq:eq2}
\end{eqnarray}
where $\bar{s}=-s$ and $N_{x,y}$ are the system sizes along the $x$- and $y$-directions, respectively. 
The spatially inhomogeneous order parameter $\Delta_{i_y,m}$ is the mean field $\Delta_{i_y,m}=-(V/N_x)\sum_{k_x}\langle c_{i_y \downarrow m}(-k_x)c_{i_y \uparrow m}(k_x)\rangle$. 

We diagonalize this Hamiltonian by the Bogoliubov transformation
\begin{eqnarray}
c_{i_ysm}(k_x)&=&\sum_{\nu}u_{i_y sm}^\nu(k_x)\gamma_\nu(k_x), \\ 
c_{i_ysm}^\dagger(-k_x)&=&\sum_{\nu}v_{i_ysm}^\nu(-k_x)\gamma_\nu(k_x).
\end{eqnarray}
The BdG equations are obtained through the commutation relation $[\mathcal{H}_{\rm MF},\gamma^\dagger_\nu(k_x)]=E_\nu(k_x) \gamma^\dagger_\nu(k_x)$. 
We numerically solve the BdG equations and calculate the order parameter and free energy of stable and metastable states
self-consistently. 
We obtain the $T$-$H$ phase diagram by comparing the free energy of different phases. 
The transition from the normal state to the superconducting state remains of the second-order for all fields. In other words, the first-order 
superconducting transition expected from the paramagnetic effect is prevented by spin-orbit coupling. 
On the basis of this observation, we determine the superconducting transition temperature 
by solving the linearized gap equation (for details, see Ref.~\citen{PhysRevB.86.134514}).

\section{Bilayer System}\label{sec:sec4}

We discuss the bilayer system with $V/t=2$ leading to the critical temperature $T_{\rm c}/t=0.0572$ for $\alpha=0$. 
In the following, we assume that the spin-orbit coupling $\alpha\gg T_{\rm c}$, as is natural for most (locally) 
non-centrosymmetric superconductors.~\cite{NCSC}

\subsection{Phase diagram}\label{ssec:ssec4-1}

\begin{figure}[htbp]
 \includegraphics[width=70mm]{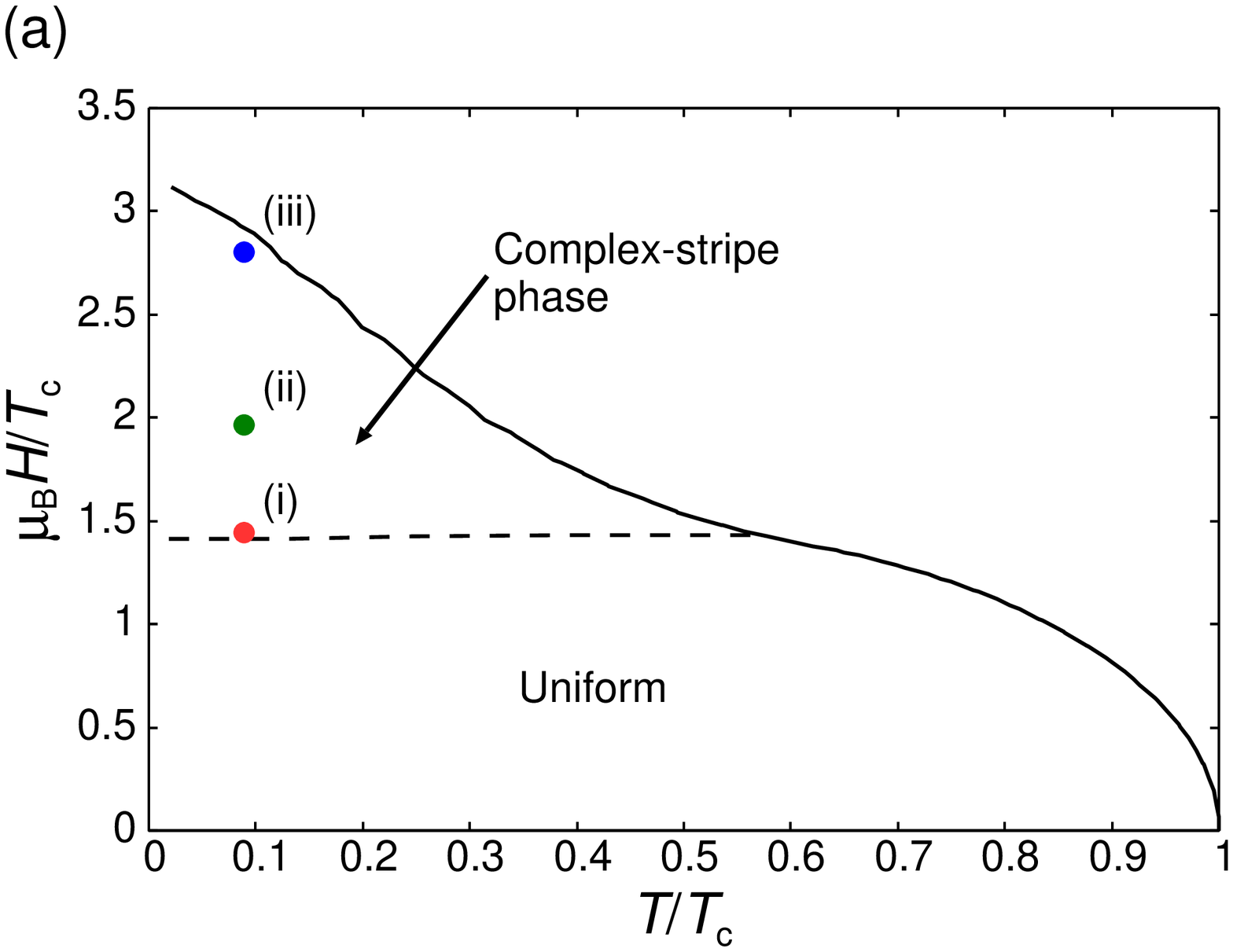}
 \includegraphics[width=70mm]{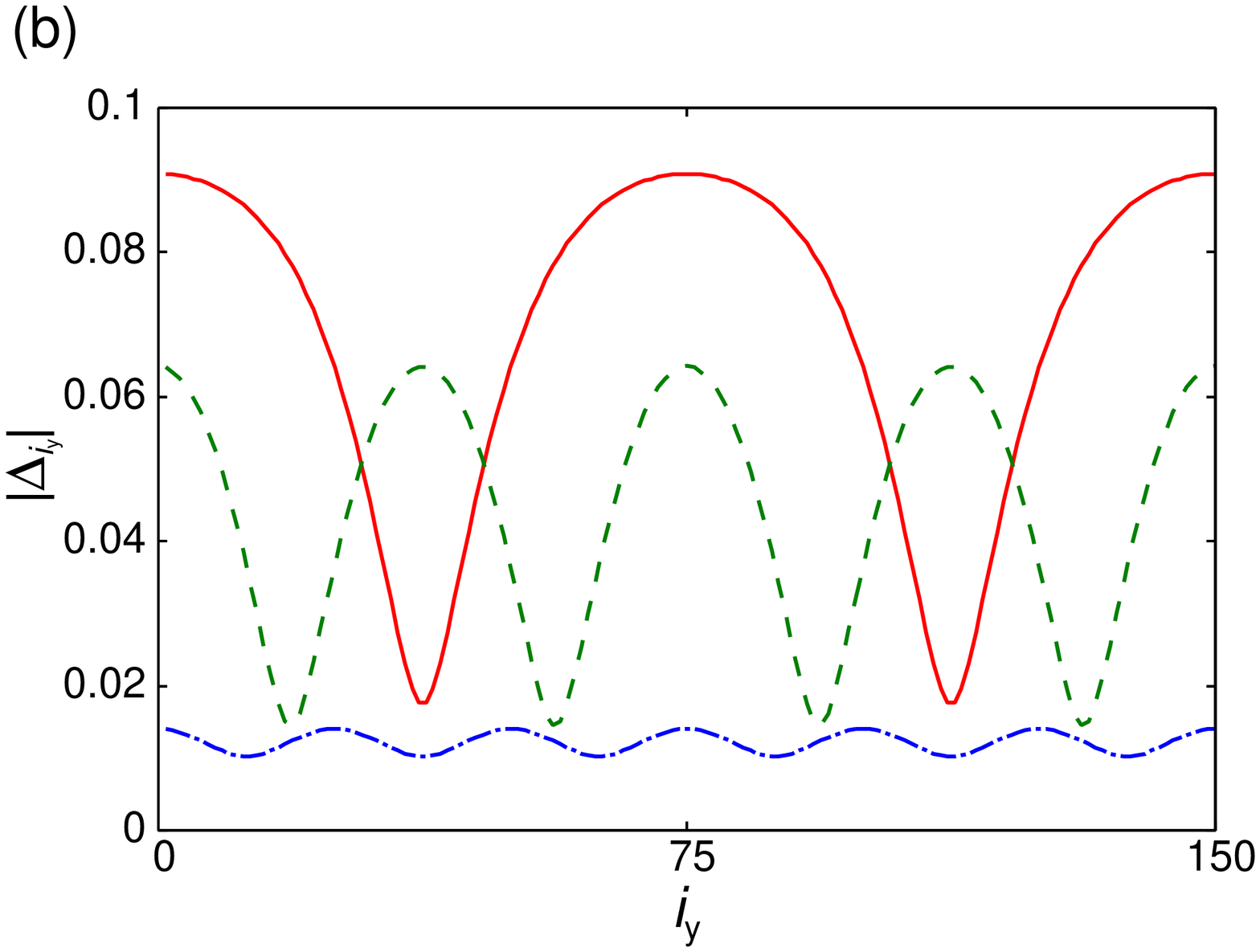}
 \caption{(Color online) (a) $T$-$H$ phase diagram for $\alpha/t_\perp=3$.
 The solid and dashed lines mark the second- and first-order phase transition lines, respectively. 
 The superconducting transition line is determined by solving the linearized BdG equation for the $500\times 500$ lattice size, and the first-order phase transition line in the superconducting state is determined by the BdG equation for the $120\times 120$ lattice size.
 The temperature $T$ and magnetic field $\mu_{\rm B}H$ are normalized by the critical temperature $T_{\rm c}=0.0560$ for $\alpha/t_\perp=3$.
 (b) Spatial profiles of the amplitude of order parameter $|\Delta_{i_y,1}|=|\Delta_{i_y,2}|=|\Delta_{i_y}|$ at (i) $(T/T_{\rm c}, \mu_{\rm B}H/T_{\rm c})=(0.0893, 1.45)$ (solid line), (ii) (0.0893, 1.94) (dashed line), and (iii) (0.0893, 2.80) (dash-dotted line). 
 We adopt the $150\times 150$ lattice size for this calculation. 
 } 
 \label{fig:fig2}
\end{figure}

The $T$-$H$ phase diagram showing the uniform and complex-stripe phase for $\alpha/t_\perp=3$ is shown in Fig.~\ref{fig:fig2}(a).
As discussed in Sect.~\ref{sec:sec2}, the complex-stripe phase is stabilized at high magnetic fields by the staggered Rashba spin-orbit coupling. 
This phase diagram is different from those of standard non-centrosymmetric superconductors~\cite{PhysRevB.75.064511,PhysRevB.76.014522} in the following two points. 
First, the common helical phase does not appear anywhere in Fig.~\ref{fig:fig2}(a), because it is suppressed by interlayer coupling, as explained in Sect.~\ref{sec:sec2}. 
Second, the complex-stripe phase appears over a wide range of the $T$-$H$ phase diagram. 
It has been found by Agterberg and Kaur that an inhomogeneous superconducting phase similar to the complex-stripe phase could be realized 
in non-centrosymmetric superconductors, but the spin-orbit coupling turns out to be rather detrimental for this type of phase.~\cite{PhysRevB.75.064511}
In contrast, the complex-stripe phase in the locally non-centrosymmetric superconductors is favored by the strong staggered antisymmetric spin-orbit coupling. 

As mentioned in Sect.~\ref{sec:sec2}, the complex-stripe phase is characterized by the spatial modulation of the
order parameter magnitude. The result of our numerical calculation is displayed in Fig.~\ref{fig:fig2}(b)
where we show profiles of $ |\Delta_{i_y,1} | $ and $ |\Delta_{i_y,2} |$ that are identical.
For this purpose, we use three sets of parameters corresponding to points (i)-(iii) in the phase diagram in Fig.~\ref{fig:fig2}(a).
For the parameter (i) $(T/T_{\rm c}, \mu_{\rm B}H/T_{\rm c})=(0.0893, 1.45)$ (solid line), 
the order parameters $\Delta_{i_y,1}$ and $\Delta_{i_y,2}$ behave similarly to that of the LO state. 
However, the order parameter has no zero nodes, but remains finite everywhere. 
With growing magnetic field the wave vector $q$ increases, as shown for the parameter (ii) [dashed line in Fig.~\ref{fig:fig2}(b)]. 
While the $q$ vector is even larger for fields close to the normal-superconductor transition, the order parameter overall shrinks and looks more
uniform and eventually disappears continuously at the phase boundary [case (iii) dashed-dotted line in Fig.~\ref{fig:fig2}(b)]. 
Note that with growing magnetic field the complex-stripe phase gradually adopts the character of an FF phase.

Let us now discuss some features of the phase diagram. The dashed line in the phase diagram of Fig.~\ref{fig:fig2}(a) is of the first-order and
is determined by the competition of the interlayer Josephson effect and the antisymmetric spin-orbit coupling. The position of this first-order transition is shifted to lower fields with decreasing interlayer coupling $t_{\perp}$. 
On the other hand, the upper critical field $ H_{\rm c2} $ is enhanced primarily by spin-orbit coupling suppressing paramagnetic depairing.~\cite{JPSJ.81.034702} The possibility to turn into a complex-stripe phase in a large part of the phase diagram 
is supported by the large $\alpha/t_\perp$, enhancing the spin-orbit coupling. 
This is in contrast to the FFLO state, which is stable only in a tiny region of the $T$-$H$ phase diagram 
in the centrosymmetric superconductors.~\cite{JPSJ.80.123706}

Note that the first-order transition from the uniform phase to the complex-stripe phase can be viewed as a lower critical field for the introduction of linear defects (``flux lines'') into interlayer Josephson junctions. 
The complex-stripe phase is induced by the paramagnetic effect in contrast to 
the conventional flux line phase due to solitons in the Josephson phase. 
The former likely occurs in the superlattice CeCoIn$_5$/YbCoIn$_5$ where a large Maki parameter 
$\sqrt{2} H_{\rm c2}^{\rm orb}/H_{\rm c2}^{\rm P}$ is observed.~\cite{NatPhys.7.849}
Then, the wavelength of the complex stripe phase $\lambda \sim \xi_{\rm ab}$ is 
much smaller than the inter-flux-line distance in the interlayer Josephson junctions 
$l \sim \Phi_0/d H \approx \xi_{\rm ab} H_{\rm c2}^{\rm orb}/H \gg  \xi_{\rm ab} $ where $ d $ is the
distance between the layers.

\subsection{Local density of states}\label{ssec:ssec4-2}

Aiming at the experimental detection of the complex-stripe phase, we now consider the local density of states (LDOS). 
With the use of our numerical solution of the BdG equation, the LDOS $\rho_{i_y}(\omega)$ can be determined straightforwardly as
\begin{eqnarray}
\rho_{i_y}(\omega)=\frac{1}{N_x}\sum_\nu\sum_{k_x}[|u_{i_y\uparrow}^\nu(k_x)|^2+|u_{i_y\downarrow}^\nu(k_x)|^2]\delta[\omega-E_\nu(k_x)],
\end{eqnarray} 
where we omit the index $m$, since the LDOS is independent of the layer for bilayer systems.  
The calculations are carried out for a lattice size of $1050\times 1050$ with use of 
a supercell technique in order to reduce the finite-size effect.

Before discussing the results of BdG equations, we illustrate the role of staggered Rashba spin-orbit 
coupling on the LDOS.  
We assume the order parameter as 
$\Delta_{i_y,1}=\Delta_{i_y,2} = \Delta_{\rm max} \cos(2\pi i_y/N_y)$ for the LO phase and $\Delta_{i_y,1}=\Delta_{i_y,2}^\ast = \Delta_{\rm max} 
[{\rm exp}({\rm i}2\pi i_y/N_y)+\delta{\rm exp}(-{\rm i}2\pi i_y/N_y) ]/(1+\delta)$ 
for the complex-stripe phase. 
We take the maximum of the order parameter as $\Delta_{\rm max}=0.0909$ and set $\delta=0.7$, leading to the LO-like complex-stripe phase.
\begin{figure}[htbp]
  \includegraphics[width=70mm]{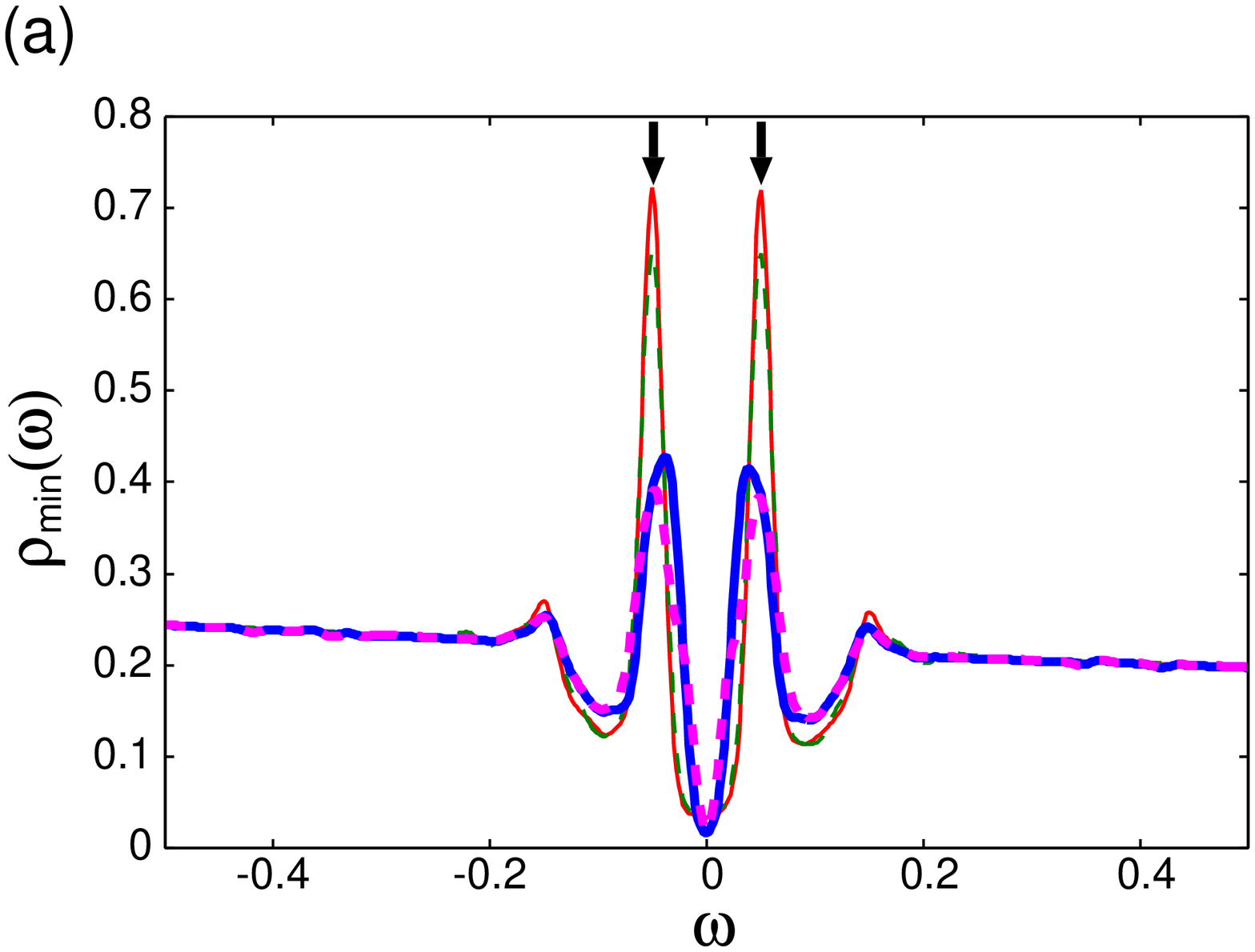}
  \includegraphics[width=70mm]{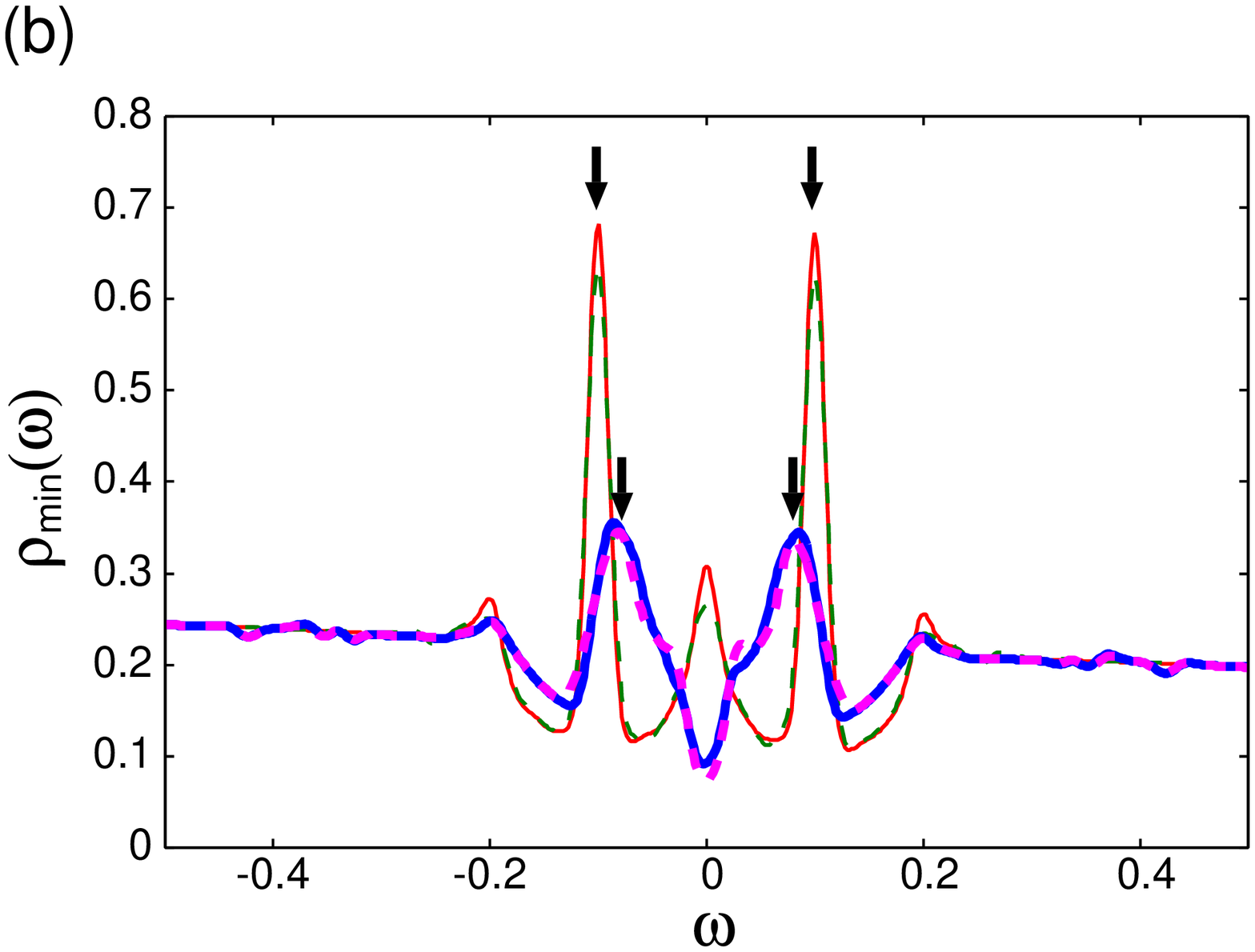}
  \includegraphics[width=70mm]{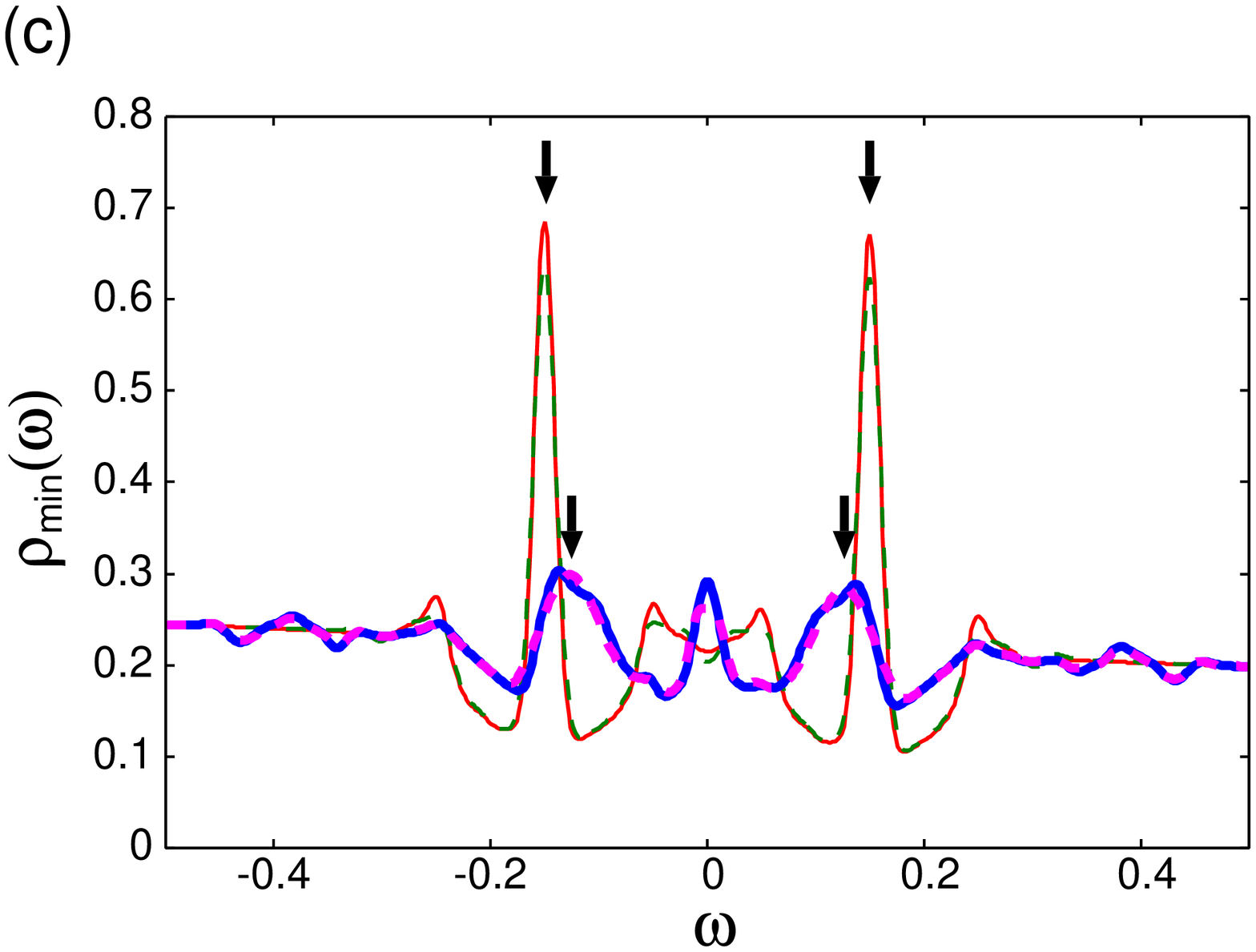}
  \caption{(Color online) LDOS $\rho_{\rm min}(\omega)$ at the position of
    minimal amplitude of the order parameter with the structure defined in the text. 
    We show the results for magnetic fields (a) $\mu_{\rm B}H=0.05$, (b) $0.1$, and (c) $0.15$. 
    The solid and dashed lines show the LDOSs for the LO and complex-stripe phases, respectively. 
    The thin and thick lines show the results for $\alpha/t_\perp =0$ and $\alpha/t_\perp=3$, respectively.
    The arrows indicate the peaks due to the (quasi-) Andreev bound states. 
    The numerical calculations are carried out for a $1050\times 1050$ lattice size using a supercell technique. 
  }
  \label{fig:fig3}
\end{figure}
Figure~\ref{fig:fig3} shows that the LDOSs are almost identical for the LO state and complex-stripe state, 
but markedly different between $\alpha/t_\perp =0$ and $\alpha/t_\perp =3$. 
In the absence of spin-orbit coupling ($\alpha/t_\perp =0$, thin lines), the Andreev bound state appears 
at $\omega=\pm \mu_{\rm B}H$.~\cite{PhysRevB.72.184501} Then, the sharp peaks in the LDOS shift with increasing 
magnetic field, as shown in Figs.~\ref{fig:fig3}(a)-\ref{fig:fig3}(c). On the other hand, in the presence of 
spin-orbit coupling ($\alpha/t_\perp =3$, thick lines), the peaks due to quasi-Andreev bound states, located at order parameter dips, 
appear in the LDOS at low fields [Fig.~\ref{fig:fig3}(a)], but then collapse with increasing magnetic field 
[Figs.~\ref{fig:fig3}(b) and \ref{fig:fig3}(c)]. 
This magnetic field dependence of the LDOS is a signature of (staggered) Rashba spin-orbit coupling and is evidence of the complex-stripe phase.

\begin{figure}[htbp]
  \includegraphics[width=70mm]{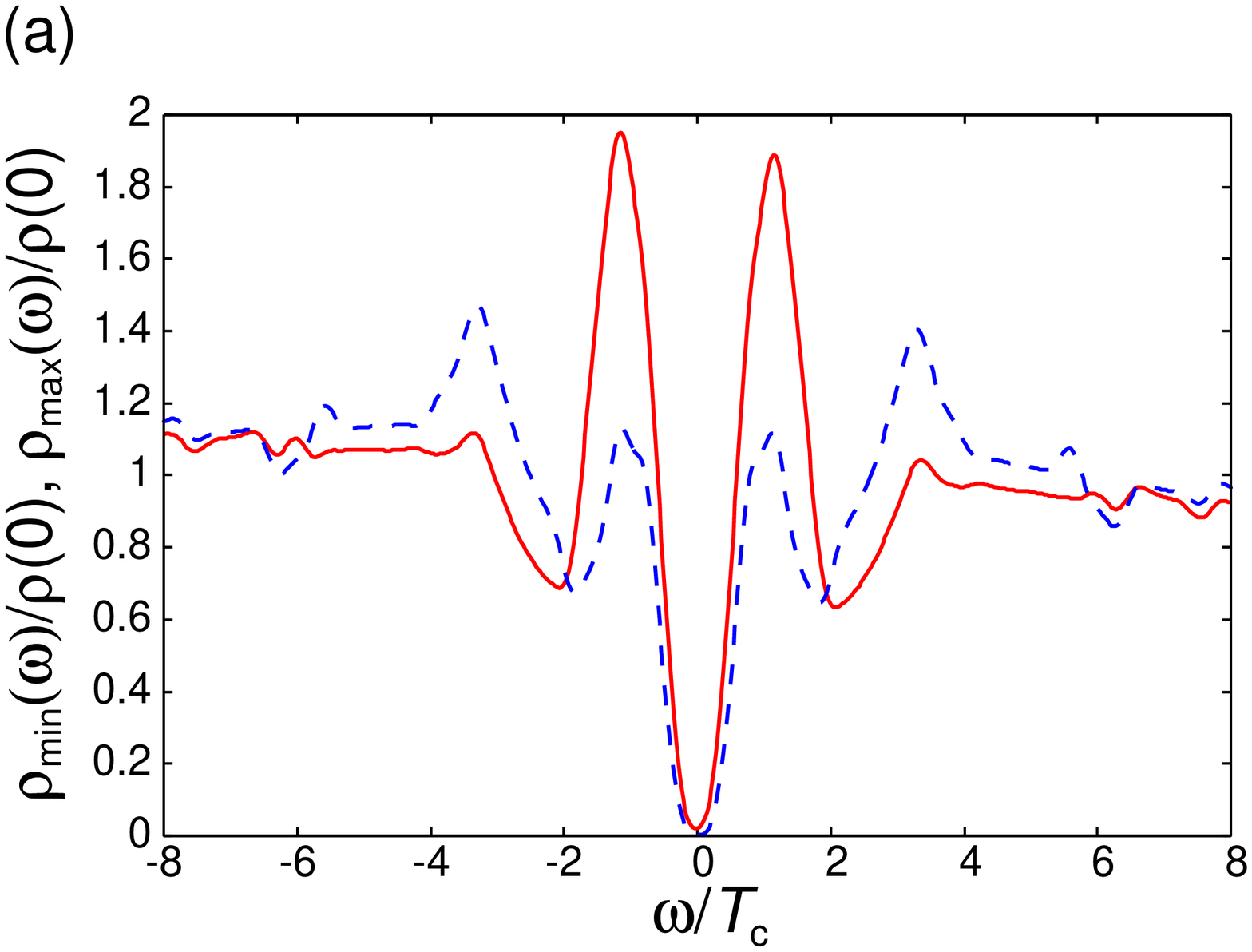}
  \includegraphics[width=70mm]{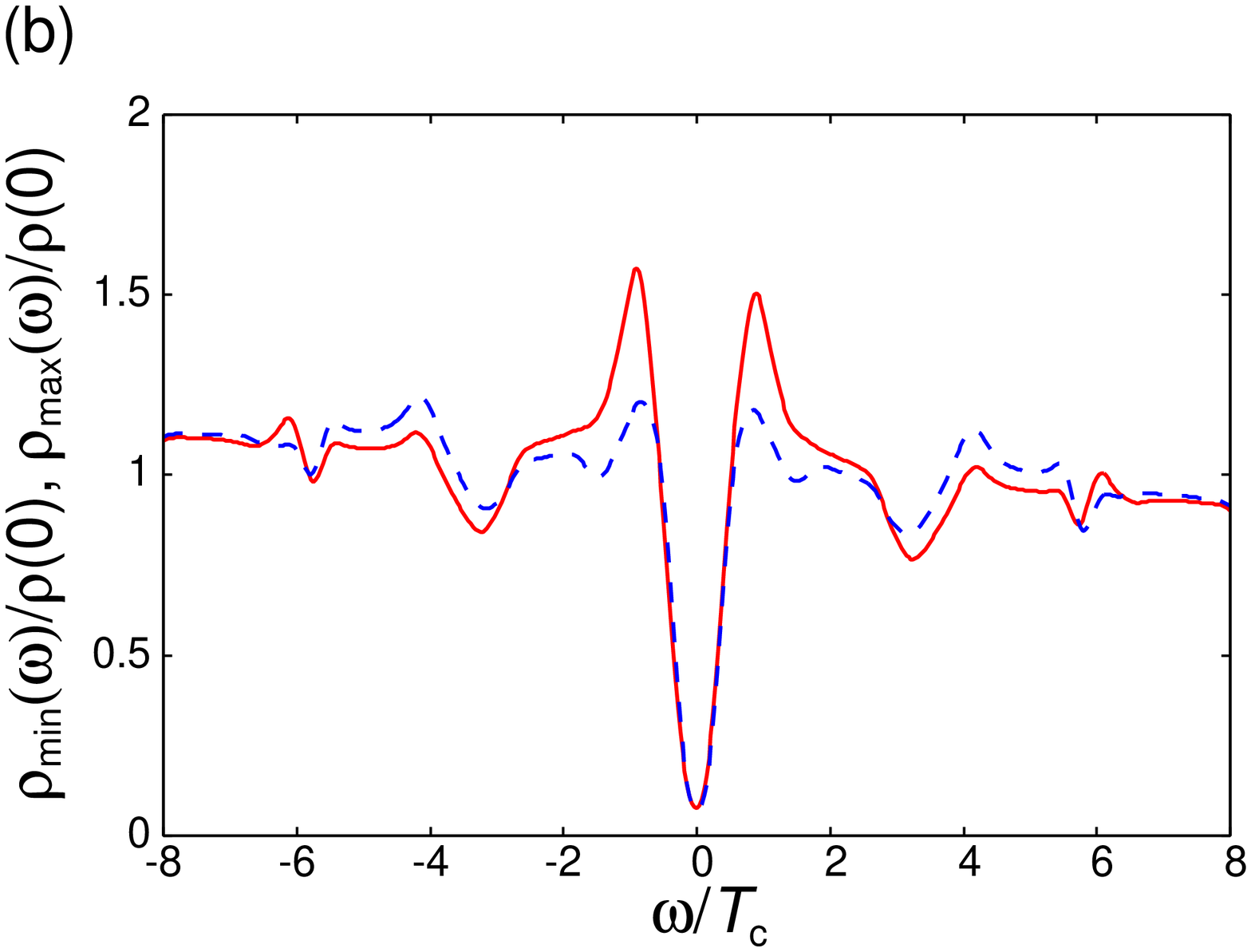}
  \includegraphics[width=70mm]{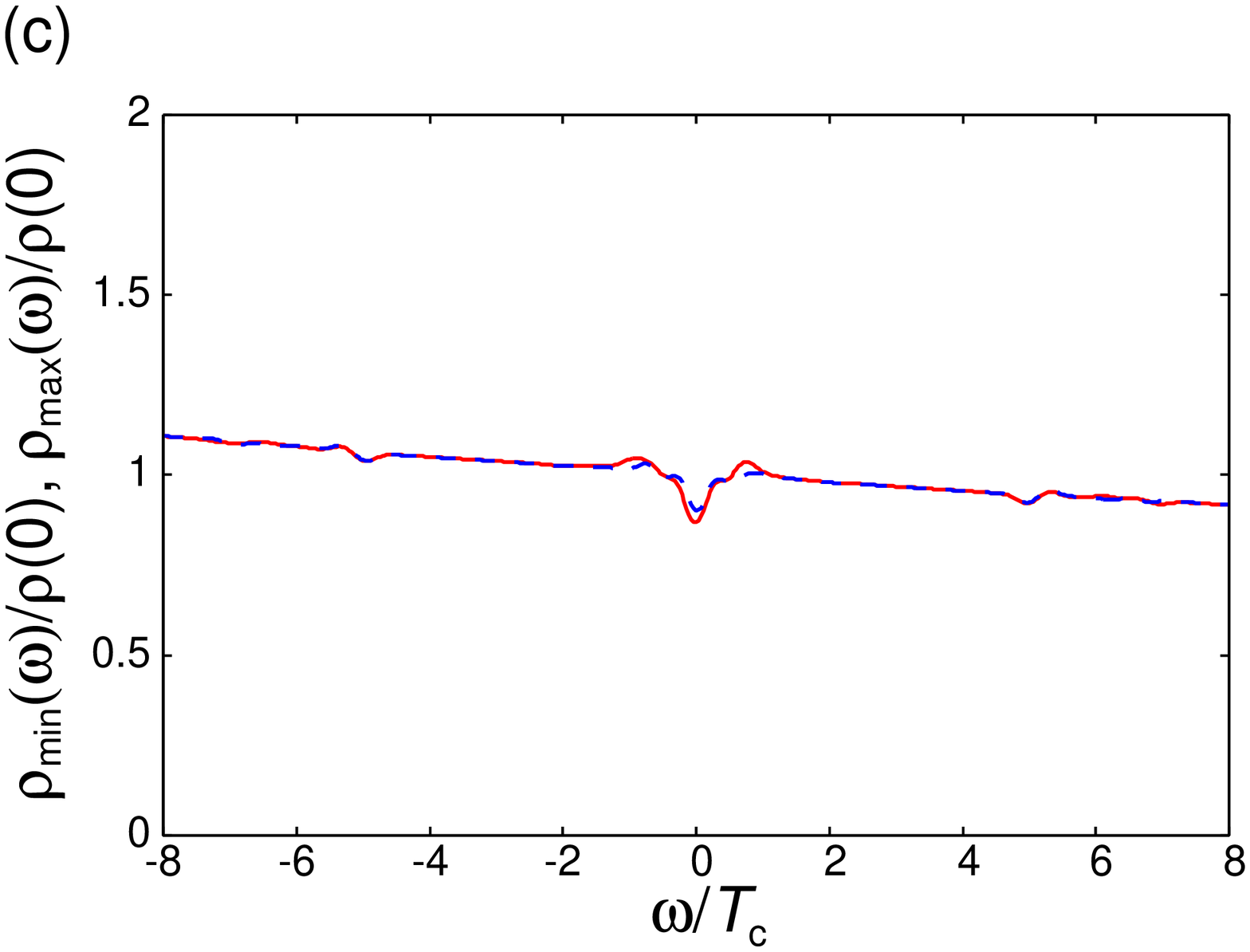}
  \caption{(Color online) LDOS calculated by solving the BdG equations. 
    We assume that $(T/T_{\rm c}, \mu_{\rm B}H/T_{\rm c})$ is equal to (a) (0.0893, 1.45), (b) (0.0893, 1.94), and (c) (0.0893, 2.80). 
    The solid lines show the LDOS $\rho_{\rm min}(\omega)$ (at the minimum of the order parameter), while dashed lines show the LDOS $\rho_{\rm max}(\omega)$ (at the maximum of the order parameter). 
    The LDOS is normalized by the density of states in the normal state, that is, $\rho(0)$. 
  }
  \label{fig:fig4}
\end{figure}

Indeed, we see the collapse of the subgap Andreev bound states in the results of the BdG equations. 
Figures~\ref{fig:fig4}(a)-\ref{fig:fig4}(c) show the LDOSs for the parameters (i), (ii), and (iii) 
in Fig.~\ref{fig:fig2}(a), respectively.
For the parameter (i) leading to the LO-like phase, the LDOS at the dips shows pronounced peaks 
at $\omega/T_{\rm c} \sim \pm 1.14$ [Fig.~\ref{fig:fig4}(a)], indicating the ``quasi-Andreev bound state'' 
in the complex-stripe phase. 
Indeed, these peaks adiabatically change to the subgap structure due to Andreev bound states  
in the LO state,~\cite{PhysRevB.72.184501} with decreasing spin-orbit coupling $\alpha$. 
However, the peaks become obscure as the magnetic field increases [Fig.~\ref{fig:fig4}(b)], as discussed above. 
Then, the gap of quasi-Andreev bound states decreases in contrast to the Andreev bound states in the 
LO state whose gap increases with the field as $\omega=\pm\mu_{\rm B}H$.~\cite{PhysRevB.72.184501}
Thus, the complex-stripe phase may be distinguished from the LO phase by investigating 
the magnetic field dependence of the LDOS using an STM experiment. 
In the high-field region near the upper critical field, the complex-stripe phase resembles the FF phase. 
Then, the spatial dependence of the LDOS vanishes and the superconducting gap becomes unclear in the LDOS [see Fig.~\ref{fig:fig4}(c)].

\section{Trilayer System}\label{sec:sec5}

\begin{figure}[htbp]
 \includegraphics[width=70mm]{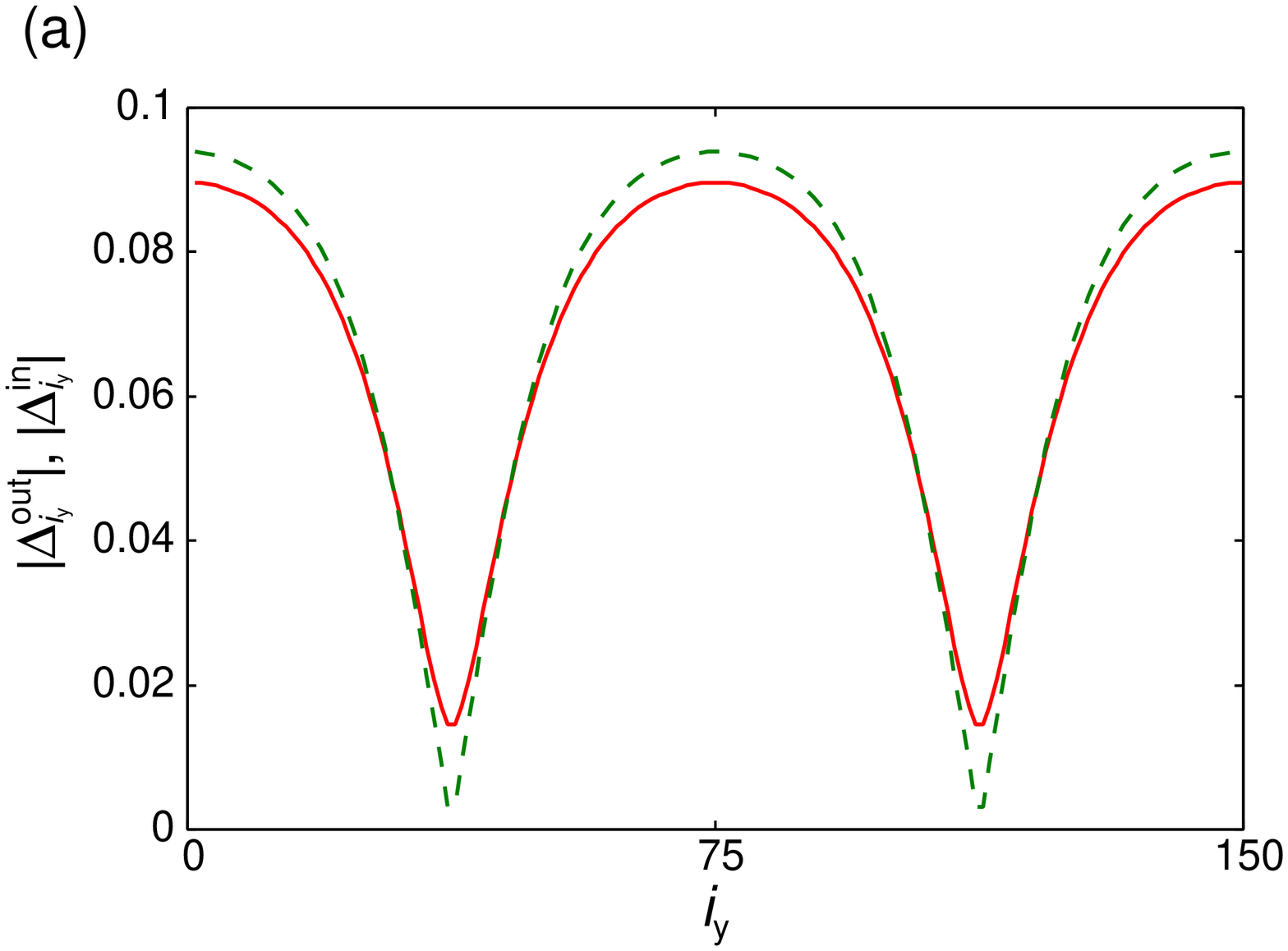}
 \includegraphics[width=70mm]{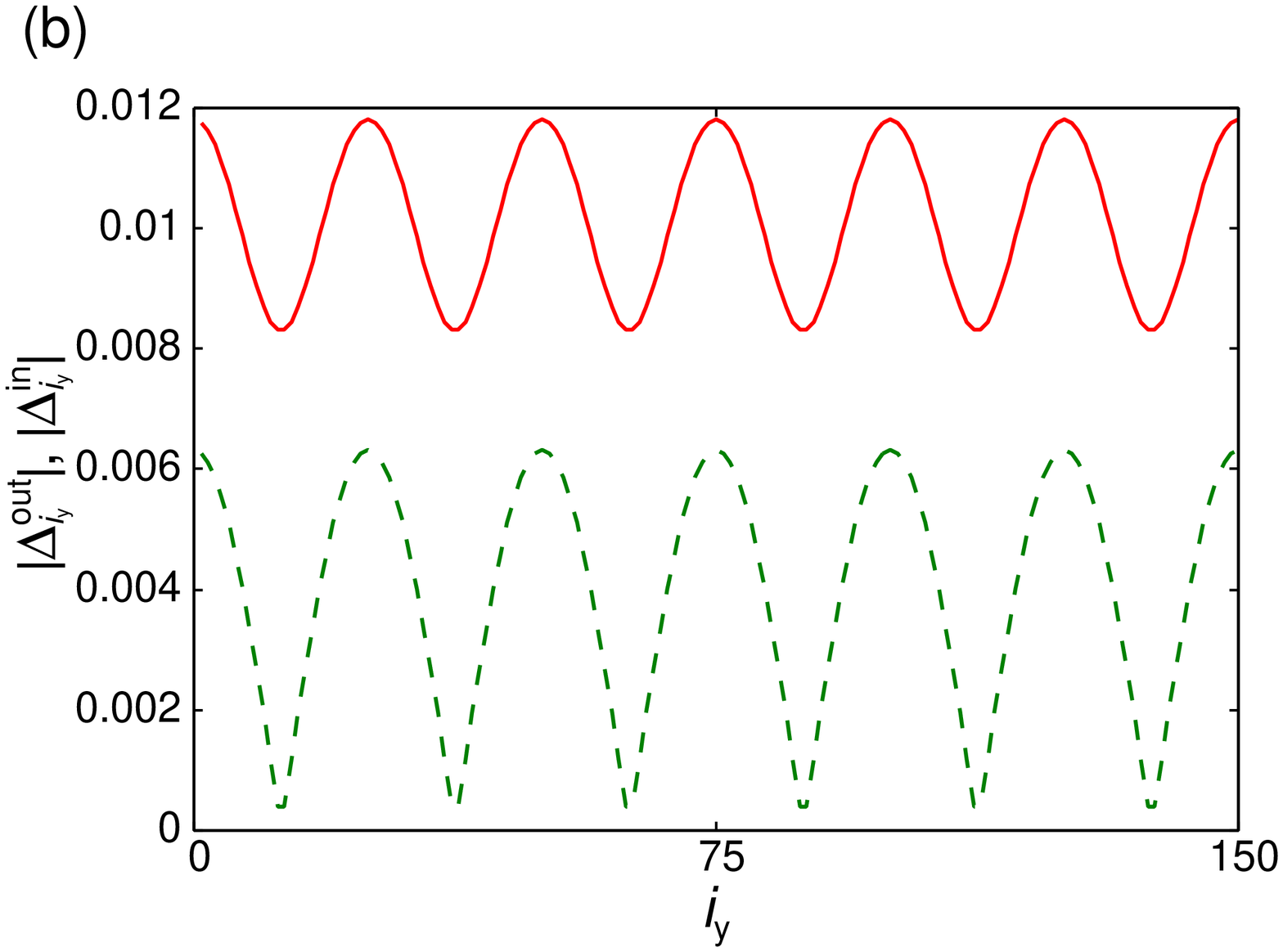}
 \caption{(Color online) Spatial profiles of the amplitude of order parameter at (a) $(T/T_{\rm c}, \mu_{\rm B}H/T_{\rm c})=(0.0885, 1.40)$ 
(complex-stripe phase near the phase boundary to the uniform phase) and 
(b) (0.0885, 2.48) (complex-stripe phase near the upper critical field), respectively. 
   The solid and dashed lines mark the amplitude of the order parameters $|\Delta_{i_y,1}|=|\Delta_{i_y,3}|=|\Delta_{i_y}^{\rm out}|$ and 
   $|\Delta_{i_y,2}|=|\Delta_{i_y}^{\rm in}|$, respectively.
   The other parameters are the same as in Fig.~\ref{fig:fig2}. 
 } 
 \label{fig:fig5}
\end{figure}

We now study the complex-stripe phase in the trilayer system. 
When we assume the attractive interaction $V/t=2$, we obtain the qualitatively same 
phase diagram as that for bilayers [Fig.~\ref{fig:fig2}(a)]. 
The order parameters of trilayer systems show, however, different features because the 
antisymmetric spin-orbit coupling vanishes in the inner layer 
as $(\alpha_1,\alpha_2,\alpha_3)=(\alpha,0,-\alpha)$. 
The outer layers have the order parameter of the complex-stripe phase, which is  
$\Delta_1({\bm r})=\Delta_0^{\rm out}({\rm e}^{{\rm i}{\bm q}\cdot{\bm r}}+\delta {\rm e}^{-{\rm i}{\bm q}\cdot{\bm r}})$ and 
$\Delta_3({\bm r})=\Delta_0^{\rm out}(\delta {\rm e}^{{\rm i}{\bm q}\cdot{\bm r}} +  {\rm e}^{-{\rm i}{\bm q}\cdot{\bm r}})$, while the 
order parameter of the inner layer is the same as that of the LO state, which is 
$\Delta_2({\bm r})=\Delta_0^{\rm in}({\rm e}^{{\rm i}{\bm q}\cdot{\bm r}} + {\rm e}^{-{\rm i}{\bm q}\cdot{\bm r}})$. 
Since the modulation vector ${\bm q}$ is along the $y$-axis as for bilayers, 
we show the amplitudes of the order parameters 
$|\Delta_{i_y}^{\rm out}| = |\Delta_{i_y,1}|=|\Delta_{i_y,3}|$
and 
$|\Delta_{i_y}^{\rm in}| = |\Delta_{i_y,2}|$
in Fig.~\ref{fig:fig5}. 
In the low-field region of the complex-stripe phase [Fig.~\ref{fig:fig5}(a)], 
the order parameters are similar for the inner and outer layers, but their zero nodes only appear on the inner layer.  
With increasing magnetic field, the order parameter for the outer layer becomes FF-like as shown 
by the small spatial dependence of the magnitude $|\Delta_{i_y}^{\rm out}|$ 
[solid line in Fig.~\ref{fig:fig5}(b)]. 
Simultaneously, the superconductivity in the inner layer is suppressed by the paramagnetic depairing effect,~\cite{JPSJ.81.034702} 
visible in the small magnitude of the order parameter $|\Delta_{i_y}^{\rm in}|$ 
[dashed line in Fig.~\ref{fig:fig5}(b)]. 
Thus, the superconductivity is essentially based on the ``helical'' order parameter of outer layers 
near the upper critical field. 

These unusual phases of the trilayer system would be strengthened by the layer dependences of the effective mass 
and charge density, which enhances the role of spin-orbit coupling.~\cite{JPSJ.82.043703} 
In contrast to that in bilayers, the complex-stripe phase in trilayers is stabilized even for a smaller $\alpha/t_\perp$,
when the imbalance of the inner and outer layers plays important roles as in multilayer 
high-$T_{\rm c}$ cuprates,~\cite{PhysRevLett.96.087001,JPSJ.80.043706} as we will show elsewhere using numerical results.

\section{Summary and Discussion}\label{sec:sec6}

In this research, we have studied the superconducting phase in a multilayer system that lacks the local inversion symmetry. 
We found that the complex-stripe phase, which is an intermediate phase between the FF and LO phases, is stabilized by the staggered Rashba spin-orbit coupling in the magnetic field.
The crossover from the FF-like phase to the LO-like phase is achieved by changing the magnetic field. 
In contrast to the helical phase in Rashba-type non-centrosymmetric superconductors, which is obscured even by the weak orbital depairing effect, the complex-stripe phase is distinguished from the vortex state because the global inversion symmetry is conserved. 
Compared with the FFLO state in centrosymmetric superconductors, the complex-stripe phase is stable in a large region of the $T$-$H$ phase diagram, because the antisymmetric spin-orbit coupling stabilizes the complex-stripe phase.

The situation discussed in this paper could be found in artificially grown superlattices of CeCoIn$_5$.~\cite{NatPhys.7.849} 
This system is regarded as the quasi-two-dimensional multilayer superconductor, and a recent measurement of the angular variation of the upper critical field showed evidence of the broken local inversion symmetry.~\cite{PhysRevLett.109.157006} 
While bulk CeCoIn$_5$ is a promising candidate for the FFLO state,~\cite{Nature.425.51,PhysRevLett.91.187004} the layer-dependent antisymmetric spin-orbit coupling supports the complex-stripe phase in the superlattice, as elucidated in this study. 
Indeed, a signature of the FFLO-like phase has been observed in the superlattice of CeCoIn$_5$.~\cite{PhysRevLett.109.157006} 
Further experimental test is desired to clarify the high-field superconducting phase of the 
superlattice CeCoIn$_5$. 

\section*{Acknowledgements}
The authors are grateful to S.~K.~Goh, D.~Maruyama, Y.~Matsuda, T.~Shibauchi, and H.~Shishido for fruitful discussions. 
This work was supported by KAKENHI (Grant Nos.~24740230 and 23102709). 
We are also grateful for the financial support from the Swiss Nationalfonds, the NCCR MaNEP, 
and the Pauli Center of ETH Zurich.

\end{document}